\documentclass[fleqn,usenatbib]{mnras}

\usepackage{newtxtext,newtxmath}

\usepackage[T1]{fontenc}

\DeclareRobustCommand{\VAN}[3]{#2}
\let\VANthebibliography\thebibliography
\def\thebibliography{\DeclareRobustCommand{\VAN}[3]{##3}\VANthebibliography}

\usepackage{graphicx}	
\usepackage{amsmath}	

\newcommand{\rate}{\mathcal R}

\title[3D Kinetic Scale Plasma Turbulence]{Electron-Scale Current Sheets and Energy Dissipation in 3D Kinetic-Scale Plasma Turbulence with Low Electron Beta}

\author[Vega et al.]{
Cristian Vega,$^{1}$\thanks{E-mail: csvega@wisc.edu}
Vadim Roytershteyn,$^{2}$
Gian Luca Delzanno$^{3}$
and Stanislav Boldyrev$^{1,2}$
\\
$^{1}$Department of Physics, University of Wisconsin-Madison, Madison, WI 53706, USA\\
$^{2}$Center for Space Plasma Physics, Space Science Institute, Boulder, Colorado 80301, USA\\
$^{3}$T-5 Applied Mathematics and Plasma Physics Group, Los Alamos National Laboratory, Los Alamos, NM 87545, USA
}

\date{Accepted XXX. Received YYY; in original form ZZZ}

\pubyear{2022}

\begin{document}
\label{firstpage}
\pagerange{\pageref{firstpage}--\pageref{lastpage}}
\maketitle

\begin{abstract}
3D kinetic-scale turbulence is studied numerically in the regime where electrons are strongly magnetized (the ratio of plasma species pressure to magnetic pressure is $\beta_e=0.1$ for electrons and $\beta_i=1$ for ions). Such a regime is relevant in the vicinity of the solar corona, the Earth's magnetosheath, and other astrophysical systems. The simulations, performed using the fluid-kinetic spectral plasma solver (SPS) code, demonstrate that the turbulent cascade in such regimes can reach scales smaller than the electron inertial scale, and results in the formation of electron-scale current sheets (ESCS). Statistical analysis of the geometrical properties of the detected ESCS is performed using an algorithm based on the medial axis transform. A typical half-thickness of the current sheets is found to be on the order of electron inertial length or below, while their half-length falls between the electron and ion inertial length.   The pressure-strain interaction, used as a measure of energy dissipation, exhibits high intermittency, with the majority of the total energy exchange occurring in current structures occupying approximately 20\% of the total volume. Some of the current sheets corresponding to the largest pressure-strain interaction are found to be associated with Alfv\'enic electron jets and magnetic configurations typical of reconnection. These reconnection candidates represent about $1$\% of all the current sheets identified. 
\end{abstract}




\section{Introduction}

Physical processes governing small-scale fluctuations in collisionless plasma turbulence are of great interest since they determine how the energy transferred down to small scales by the turbulent cascade is dissipated into the heating and acceleration of various plasma species. It is generally understood that at scales below inertial scales and/or gyro-scales of ions their motion decouples from fluctuations of the magnetic field, leading to the appearance of kinetic- and inertial-Alfv\'en modes that may experience collisionless damping and hence dissipate turbulent energy \cite[e.g.,][]{howes08a,schekochihin09,boldyrev12b,alexandrova2012,boldyrev2015,franci2015,told2015,chen2016,cerri2017a,groselj2018,verscharen2019,arzamasskiy2019,zhou2022}.

A universal property of turbulence in magnetized plasmas is also the emergence of intermittency of magnetic field fluctuations \cite[e.g.,][]{Karimabadi2013,zhdankin2014,papini2021}. This manifests itself in the appearance of current sheets, which are narrow structures with intense shear of a magnetic field. The current sheets enable efficient energy transfer between fields and particles, and may undergo tearing instability and magnetic reconnection \cite[e.g.,][]{matthaeus_turbulent_1986,wan2012,Karimabadi2013,zhdankin_etal2013,wan2013,wan2015,zhdankin2012a,zhdankin2016a,zhdankin2016b,cerri2017b,franci2017,Haggerty2017,loureiro2017,mallet_disruption_2017,walker2018,dong2018,camporeale2018,comisso2018,boldyrev2019,papini2019,vega2020}. Quantifying the relative contribution of the current sheets to the average turbulence dissipation rate is of considerable interest.

The relative strength and significance of various processes operating at sub-proton scales vary with plasma parameters. Plasma systems where the parameter $\beta_e$, the ratio of electron thermal pressure to the magnetic pressure, is significantly less than one can exhibit particularly rich small-scale dynamics because in such cases a new range of scales between the electron inertial scale $d_e$ and the electron gyroradius $\rho_e$ becomes accessible to plasma fluctuations.\footnote{Recall that for particle species $\alpha$, $\beta_\alpha=\rho_\alpha^2/d_\alpha^2$} In particular, electron-scale current sheets are expected to be a prominent feature of such turbulent regimes ~\citep{chen_boldyrev2017,roytershteyn2019,vega2020}. This opens the possibility that dissipation associated with such current sheets and, possibly, with magnetic reconnection, play a significant role in the overall energy dissipation in low-$\beta_e$ turbulence.
Recent observations of sub-ion scale turbulence in the Earth's magnetosheath gave support to this expectation by demonstrating the possibility of magnetic reconnection in electron-scale current sheets, which does not involve coupling to the ions ~\citep[]{phan2018}. 

The properties of such ``electron-only" magnetic reconnection have been investigated using numerical simulations and theoretical considerations \cite[e.g.,][]{phan2018,pyakurel2019,vega2020,pyakurel2021,franci2022}. It has been suggested that the length of the current sheet along the electron outflow is a key property that differentiates between the onset of ``electron-only" and traditional reconnection that involves coupling to ions \cite[][]{pyakurel2019,vega2020,stawarz2022}. It is natural to expect that in the low-$\beta_e$ regimes, current sheets are formed by turbulence with small enough scales that favor the ``electron-only" reconnection. Previous numerical simulations indeed demonstrated the existence of electron-only reconnection events in low-$\beta_e$ turbulence~\citep{vega2020} but did not attempt to estimate the overall significance of energy dissipation at electron-scale current sheets.

Numerical simulations of low-$\beta_e$ turbulence, especially fully kinetic simulations, are associated with significant computational challenges due to increased separation between various characteristic plasma scales (inertial length, particle gyro-radius, Debye length, etc). As a result, many of the previous studies of turbulence in the regime of interest to this work had a number of important limitations.

In particular, most were two-dimensional (2D) and/or  assumed relatively large perturbations  of the magnetic field at kinetic scales $\delta B/B_0 \sim  1$ in order to overcome numerical noise inherent to frequently adopted particle-in-cell techniques ~\cite[e.g.,][]{parashar2018,adhikari2021,franci2022,roy2022}. The latter condition may be appropriate for situations where the scale at which the energy is supplied to turbulent fluctuations is not far removed from the kinetic scales (as is the case in the Earth's magnetosheath). But in other systems where low-$\beta_e$ turbulence exists (e.g., at small heliospheric distances or in the outer solar corona), the magnitude of the magnetic fluctuations at kinetic scales is very small $\delta B/B_0 \ll 1$ \cite[e.g.,][]{chen2020,kasper2021}.

In this paper, we analyze a 3D numerical simulation of kinetic scale turbulence in the low-$\beta_e$ regime, $\beta_e=0.1$, $\beta_i=1$. The simulation was performed using the SpectralPlasmaSolver (SPS) code~\citep{Vencels2016,roytershteyn2018}, which is proving to be promising in studying this plasma turbulence regime \citep{roytershteyn2019}. SPS utilizes a spectral representation of the distribution function in the velocity space that, when truncated to a finite (small) number of terms, results in a hierarchy of models intermediate between conventional two-fluid and kinetic descriptions. Crucially, even with a limited number of basis functions, the SPS algorithm provides a suitable approximation for collisionless damping of low-frequency fluctuations, while also capturing physics inaccessible to other reduced models (e.g., high-frequency fluctuations ordered out of gyrokinetic codes; see, for example, ~\cite[]{roytershteyn2018}). Further, SPS utilizes a fully implicit, noise-free algorithm~\citep{delzanno2015} that allows it to avoid many of the limitations of traditional fully kinetic codes such as explicit particle-in-cell (PIC) codes. For instance, it allows one to perform simulations with larger ratios of plasma frequency to cyclotron frequency (and thus better ensure plasma quasineutrality), and to address turbulent regimes with smaller magnetic fluctuations with respect to the guide field.

One of the major goals of this work is to examine the contribution of electron-scale current sheets to particle energization and discuss the question of to what extent particle energy conversion can be associated with magnetic reconnection. To achieve this, we perform a statistical analysis of the electron-scale current sheets generated in 3D low-$\beta_e$ turbulence using the algorithm based on a medial axis transform borrowed from image processing~\cite[][]{Blum1967}. This method allows for a careful analysis of current sheets with complicated shapes. We discuss the distribution of their sizes and the associated energy dissipation rates. We observe that electron-scale current sheets contribute significantly to particle energization as measured by the pressure-strain interaction.

As discussed above, the interplay between magnetic reconnection and turbulence is a question of great interest. However, it is generally difficult to identify and study reconnection sites in 3D. In previous studies, reconnection sites were found in 3D simulations by identifying sites where numerous signatures of reconnection were present simultaneously (e.g., large current density, particle heating, fast ions and electrons; see, for example, \cite[]{Rueda2021}). In this work, we take a similar approach and find candidates for electron-only reconnection by searching for large values of the pressure-strain interaction (a signature of particle heating~\citep{Yang2017,yang2022}) and large variations in the electron fluid velocity (a signature of fast electron outflows) in the vicinity of strong current peaks. A close inspection of each candidate site picked up by these criteria reveals, however, that not all of them are, in fact, good reconnection candidates. As an illustration, we discuss in detail the structure of one such electron-only reconnection event detected in our simulations.

In what follows we will use the following definitions: $\omega_{p\alpha}=(4\pi n_0e^2/m_\alpha)^{1/2}$ and $\Omega_{c\alpha}=|e|B_0/(cm_\alpha)$ are the plasma and cyclotron frequencies for species $\alpha$ with mass $m_\alpha$, charge $e$, and average particle density $n_0$, in background field $B_0$. The ratio of plasma to magnetic pressure is $\beta_\alpha=8\pi n_0T_\alpha/B_0^2$. The inertial length is $d_\alpha=c/\omega_{p\alpha}$ and particle gyro-radius is $\rho_\alpha = d_\alpha \sqrt{\beta_\alpha}$.

\section{Numerical setup}

The SpectralPlasmaSolver (SPS) code was used to perform a 3D simulation of sub-ion-scale plasma turbulence. SPS solves the kinetic equations for all plasma species by expanding the plasma distribution function in Hermite functions (akin to a moment expansion). It results in a truncated set of three-dimensional partial differential equations (PDEs) for the expansion coefficients whose expressions can be explicitly found in Refs.~\cite{delzanno2015,roytershteyn2018}. In principle, the number of expansion terms can be arbitrary and one can treat the closure of the truncated system parametrically. In practice, the available computer power limits the number of expansion terms that can be used for a given problem. SPS solves the PDEs via a Fourier discretization in physical space and uses an implicit (midpoint rule) time discretization. The discrete equations are then solved with a Jacobian-Free Newton-Krylov approach with preconditioning, as detailed in Refs.~\cite{delzanno2015, Vencels2016, roytershteyn2018}.  An artificial collisional operator~\cite{delzanno2015} is used to damp the higher-order coefficients and to prevent the well-known recurrence phenomenon typical of high-order methods for the solution of the Vlasov equation. By construction, the collision operator conserves total mass, momentum, and energy and is applied directly only on the higher-order moments.

The simulation was initialized with a uniform two-species plasma with Maxwellian velocity distributions corresponding to $\beta_e=0.1$, $\beta_i=1$, and the same uniform density $n_0$ for each species, embedded into a uniform magnetic field of strength $B_0$ oriented in the $z$ direction. The ratio of the plasma electron frequency to the electron cyclotron frequency was \(\omega_{pe}^2/\Omega_{ce}^2=10^4\), and the ion-to-electron mass ratio was  \(m_i/m_e=100\). The dimensions of the simulation domain were $L_x=L_y=10d_i=100d_e$ and $L_z=60d_i=600d_e$, and periodic boundary conditions were used. The spatial domain was decomposed into $511\times511\times63$ Fourier modes and the velocity domain in $4\times4\times4$ Hermite modes. The reduced spatial resolution in the z-direction reflects the fact that in the presence of a strong guide field, turbulent fluctuations are anisotropic, approximately satisfying the critical balance condition, $k_z/k_\perp \sim \delta B/B_0$ \cite[e.g.,][]{boldyrev12b,tenbarge2012,boldyrev2021}. Clearly, the use of reduced resolution in the parallel direction excludes the possibility of describing fluctuations with very short parallel wavelengths. At present, there is little evidence to suggest that coupling to such fluctuations is an important process in the turbulent regime considered. Consequently, the resolution was chosen to reduce the numerical cost of the simulation in this first study. The model utilized here can be thought of as an advanced two-fluid model retaining 128 fluid moments (64 per species) and utilizing a closure at the level of heat flux tensor~\citep[see, e.g., discussion in][]{delzanno2015,roytershteyn2018}. The artificial collisional operator had collisionality $\nu = 0.01 \omega_{pe}$~\citep{delzanno2015}. The time step was $\Omega_{ce}\delta t=1$.  We emphasize that the implicit time discretization used in SPS allows us to study a regime with rather large $\omega_{pe}^2/\Omega_{ce}^2$, which is extremely challenging for algorithms based on explicit time discretization, such as those used in many production PIC codes.

Decaying turbulence was seeded by imposing randomly phased initial perturbations of the magnetic and velocity fields of the type 
\begin{eqnarray}
\delta\mathbf{B}=\sum_k\delta\mathbf{B}_k\cos(\mathbf{k}\cdot\mathbf{x}+\chi_k),\\
\delta\mathbf{V}=\sum_k\delta\mathbf{V}_k\cos(\mathbf{k}\cdot\mathbf{x}+\phi_k), 
\end{eqnarray}
with the wave numbers $\mathbf{k}=\{2\pi l/L_x,\, 2\pi m/L_y,\, 2\pi n/L_z\}$, where $l,m=-2,...,2$ and $n=0,...,2$. Since we initialize $k_z = 0$ modes, certain modes are included in the sum twice (e.g., $(k_x,k_y,0)$ and $(-k_x, -k_y, 0)$), so the sum is further restricted to make sure that each $\mathbf{k}$ is sampled only once. The injection scale of turbulence is thus $k_\perp d_i\approx2$. The amplitudes of the initial modes satisfy conditions $\mathbf{k}\cdot\delta\mathbf{B}_k=0$, $\mathbf{B}_0\cdot\delta\mathbf{B}_k=0$, $\mathbf{k}\cdot\delta\mathbf{V}_k=0$, and $|\delta\mathbf{B}_k|/B_0=|\delta\mathbf{V}_k|/V_A$, where $V_A=B_0/(4\pi n_0m_i)^{1/2}$ is the Alfvén speed. The initial root-mean-square fluctuation was $\langle\delta B^2(\mathbf{x},t=0)\rangle^{1/2}/B_0=\langle\delta V^2(\mathbf{x},t=0)\rangle^{1/2}/V_A\approx0.071$.\\


\section{Numerical results}
\begin{figure}
\includegraphics[width=1\columnwidth,height=0.76\columnwidth]{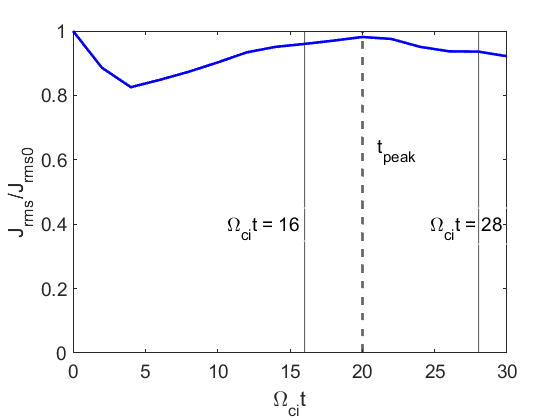}
\includegraphics[width=1\columnwidth,height=0.76\columnwidth]{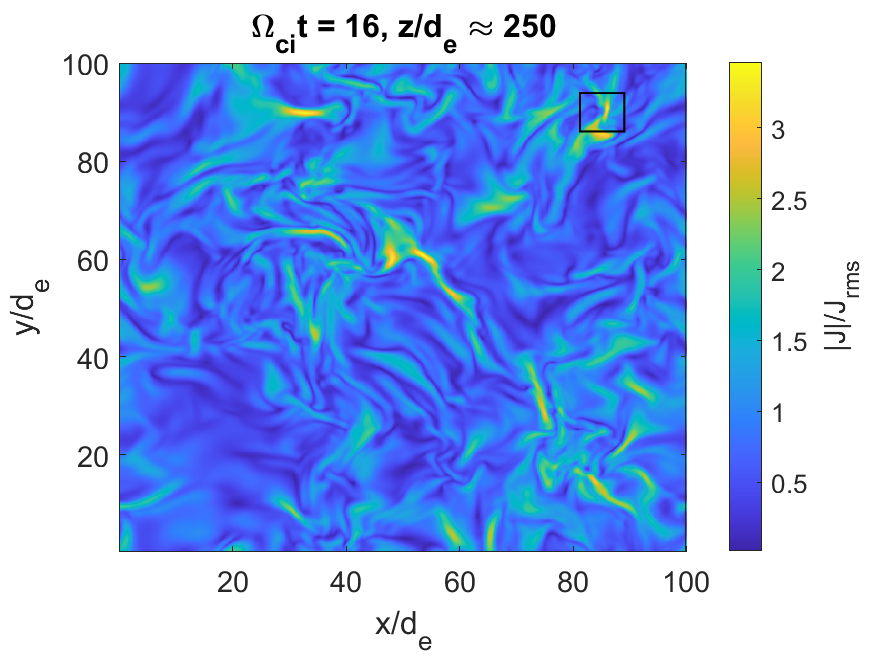}
\includegraphics[width=1\columnwidth,height=0.76\columnwidth]{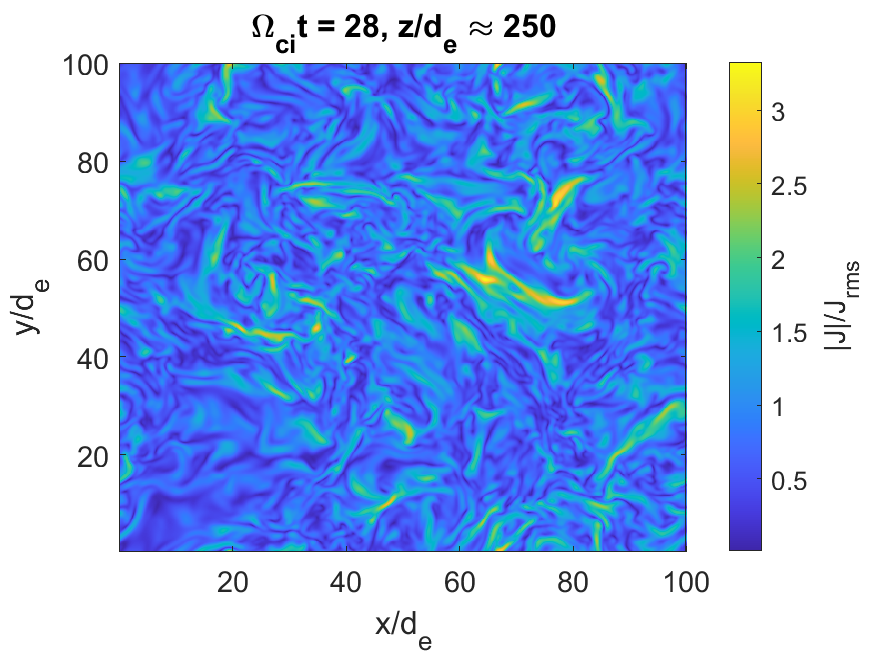}
\caption{Top panel: Time evolution of the root-mean-square current density $J_{\text{rms}}$. At $\Omega_{ci}t=20$ (dashed line) $J_{\text{rms}}$ starts to decrease. Note that from $\Omega_{ci}t=16$ to $\Omega_{ci}t=28$, $J_{\text{rms}}$ fluctuates only by about 5\%. Middle panel: Mean-field-perpendicular cut of the current density $|\mathbf{J}|$ shortly before a peak in $J_{rms}$ is reached. The rectangle on the top right encloses a reconnection site (see discussion on section \ref{Rec}). Bottom panel: Mean-field-perpendicular cut of the current density $|\mathbf{J}|$ after a peak in $J_{rms}$ is reached. More current sheets are visible relative to the middle panel.}
\label{Jrms}
\end{figure}

The top panel of Figure~\ref{Jrms} shows the time evolution of the root-mean-square current density $J_{\text{rms}}$, normalized to $J_{\text{rms0}}=J_{\text{rms}}(t=0)$. After decaying for about four gyration times from an initial maximum, $J_{\text{rms}}$ reaches another local maximum at $\Omega_{ci}t=20$ (shown with a dotted vertical line labeled $t_{\text{peak}}$) and decays thereafter. We note that $\Omega_{ci}t\sim20$  is close to Alfv\'en crossing time $\tau_c = \ell_z/V_A$, with $\ell_z=L_z/2$. The two solid vertical lines correspond to the times that are analyzed in this section: we study in detail an electron-only reconnection site found at $\Omega_{ci}t=16$ and we discuss the statistics of current sheets and energy dissipation at $\Omega_{ci}t=28$. Both time slices illustrate dynamical phenomena that are important in developed turbulence.

The middle and bottom panels of Figure~\ref{Jrms} show a mean-field-perpendicular cut of the current density $|\mathbf{J}|$ at $\Omega_{ci}t=16$ and $\Omega_{ci}t=28$, respectively. The reconnection site to be analyzed in section \ref{Rec} can be seen on the top-right corner of the middle panel, enclosed by a black rectangle. While $J_{\text{rms}}$ fluctuates only by about 5\% from $\Omega_{ci}t=16$ to $\Omega_{ci}t=28$, more small-scale structures, including sheet-like structures, can be seen at the later time.

\subsection{Power Spectra and Compressibility}

\begin{figure}
\includegraphics[width=1\columnwidth,height=0.76\columnwidth]{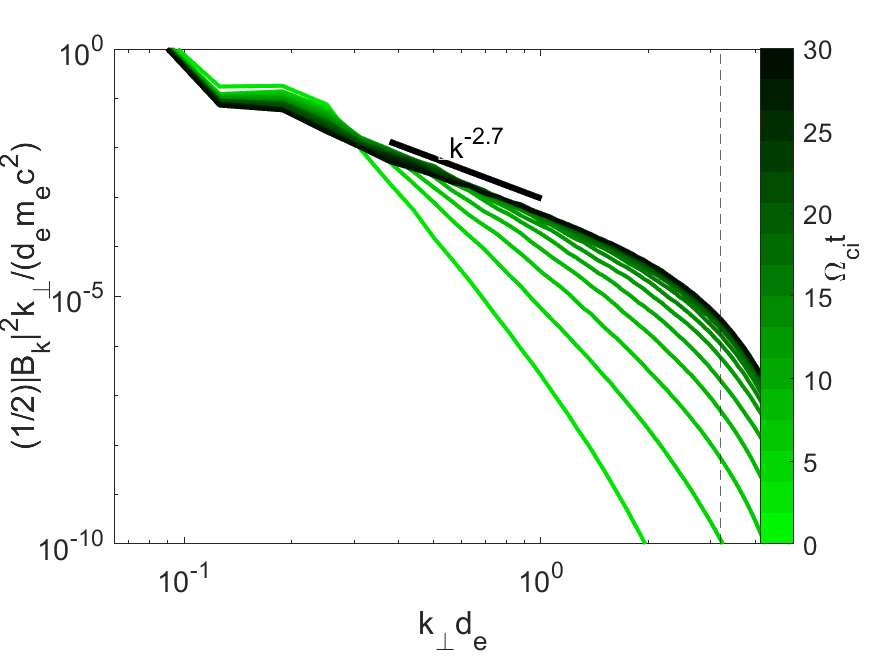}
\includegraphics[width=1\columnwidth,height=0.76\columnwidth]{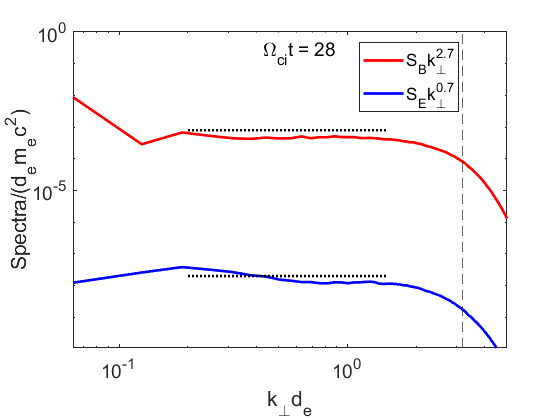}
\includegraphics[width=1\columnwidth,height=0.76\columnwidth]{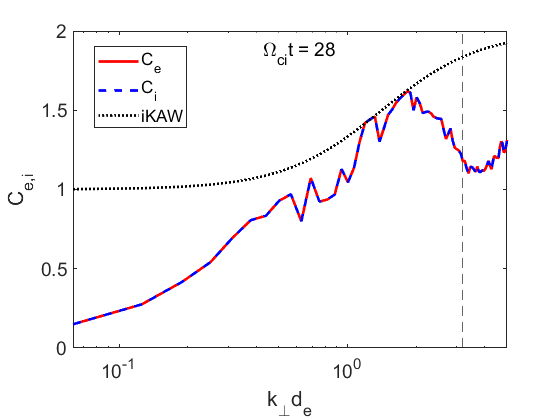}
\caption{Top panel: Time evolution of magnetic energy spectrum. Middle panel: Compensated electric (blue) and magnetic (red) energy spectra. Bottom panel: Electron (red) and ion (blue) compressibilities. The black dotted line shows the analytic prediction for the inertial kinetic-Alfv\'en modes. The vertical dashed line shows the electron gyroscale.}
\label{spectra}
\end{figure}

The top panel of Figure~\ref{spectra} shows the time evolution of the magnetic energy spectrum. It approaches a power-law spectrum close to $k_\perp^{-2.7}$ by $\Omega_{ci}t\sim20$.
The middle panel shows the compensated magnetic and electric energy spectra, $S_B(k_\perp)k_\perp^{2.7}$ and $S_E(k_\perp)k_\perp^{0.7}$, respectively, at time $\Omega_{ci}t=28$. The energy spectra $S_{B,E}$ were computed with the 2D Fourier decomposition of the electric and magnetic fields in the plane perpendicular to the mean magnetic field. The Fourier spectrum of the fields was used to find the energy spectrum on a given plane (with background field-parallel coordinate $z$) and a given direction (with polar coordinate $\phi$). The energy spectra $S_{B,E}$ were then obtained as the average over both coordinates (e.g., $S_B=\langle|B_{k_\perp}(\phi,z)|^22\pi k_\perp\rangle_{\phi,z}$, where $\langle...\rangle_{\phi,z}$ is the average over $\phi$ and $z$). 
The observed power law behavior in the range $0.1<k_\perp d_e<1$ is consistent with the observations of sub-ion-scale turbulence in the solar wind upstream of the Earth’s bow shock ~\cite[e.g.,][]{bruno2013,chen2016} and in the previous simulations ~\cite[e.g.,][]{roytershteyn2019}. We note that the value of $\beta_e=0.1$ is not low enough to enable a significant separation of scales between the electron gyroradius and the electron inertial length. Consequently, the asymptotic inertial kinetic Alfvén turbulence regime~\citep{chen_boldyrev2017,roytershteyn2019}, if present in the simulation, is not fully realized. The spectra exhibit a sharp steepening at $k_\perp\rho_e\sim1$, which is consistent with the expected onset of damping at those scales and the typical thickness of the current sheets (see below).

The bottom panel of Figure~\ref{spectra} shows the electron and ion compressibilities $C_{e,i}=(\delta n_{e,i}/n_0)^2/(|\delta\mathbf{B}|/B_0)^2$ compared against analytical predictions for low-frequency inertial kinetic Alfv\'en waves in \citet{chen_boldyrev2017}. There exists a range of scales around $k_\perp d_e \sim 1$ where the simulation results seem to approach the analytical predictions. This is consistent with the result that turbulent fluctuations retain certain properties of linear waves even in a strongly nonlinear system \cite[e.g.,][]{maron_g01,tenbarge2012}. Also notable is the fact that, in contrast to the case with a low value of the so-called quasineutrality parameter $\omega_{pe}^2/\Omega_{ce}^2$ presented in~\citet{roytershteyn2019}, the plasma  quasi-neutrality is well maintained at all scales in the presented simulations, since the quasineutrality parameter is sufficiently large here,   $\omega_{pe}^2/\Omega_{ce}^2=10^4$.

\subsection{Statistics of current sheets}

\begin{figure}
\hskip-8mm\includegraphics[width=1.2\columnwidth,height=0.86\columnwidth]{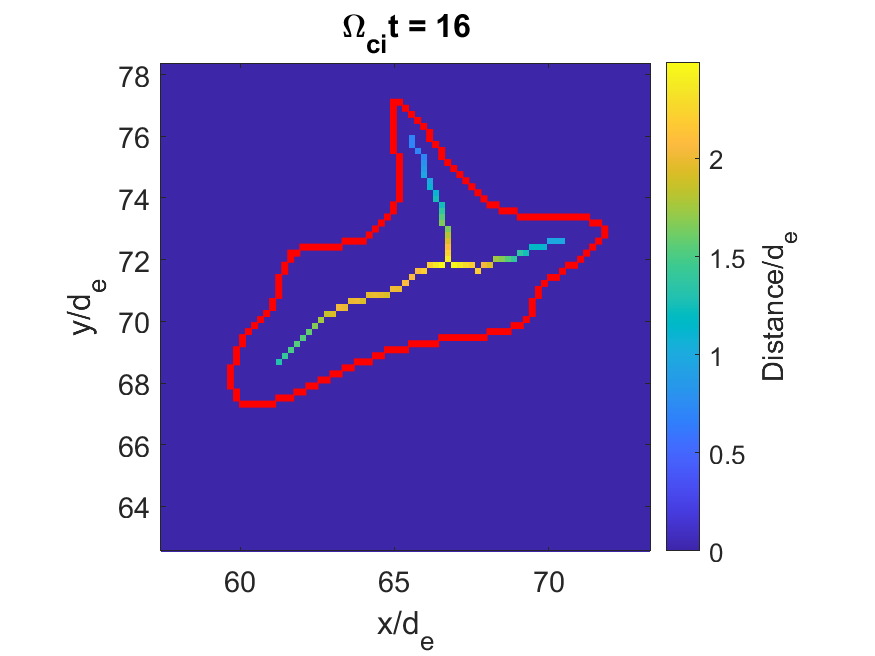}

\hskip-8mm\includegraphics[width=1.2\columnwidth,height=0.86\columnwidth]{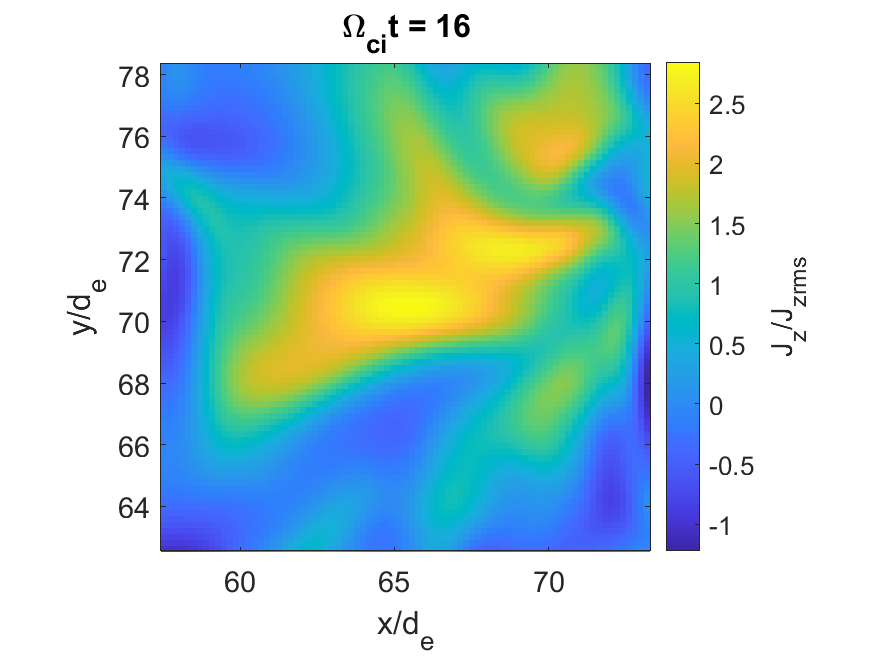}
\caption{Top panel: Example of skeletonization of a current sheet. The red curve traces the boundary of the current sheet as defined in the text. The curve within is the skeleton obtained with the medial axis transform, with the color code representing the shortest distance to the red boundary. Bottom panel: Color map of the same current sheet.}
\label{Jsize}
\end{figure}

A major focus of the present work is on characterizing the properties of small-scale current sheets, and on quantifying their contribution to energy dissipation. Numerically, we identify the current sheets by using the same algorithm that was used in \cite{vega2020}. First, the algorithm finds current peaks with a current density greater than twice the root-mean-square (rms) value and then looks for points around the peak where the current density is at least 50\% of the peak. The algorithm is applied on all 2D planes perpendicular to the uniform  background magnetic field ($z$-direction). 
This way, 2D current sheets are identified. (Obviously, several 2D current sheets found in such a procedure may belong to the same 3D structure.) In what follows, we study the statistics of such 2D current sheets.

{In order to define the half-thickness of a 2D current sheet, we first apply the so-called medial axis transform to the current sheet. Such a transform was originally introduced in~\citet{Blum1967}. It associates the so-called skeleton (or medial axis) with the current sheet boundary. The skeleton is defined as the locus of points that have more than one closest point on the boundary of the object. In 2D, the skeleton consists of the centers of circles that are tangent to the contour at at least two points. The skeleton preserves the topology of the original shape so that the method is well-suited for defining the thickness of current sheets with complex shapes.  Figure \ref{Jsize} shows an example of the ``skeletonization" of a 2D current sheet. The boundary of the current sheet is shown in red and the skeleton is shown in a color map, which represents the distance of each point of the skeleton to the boundary (that is, the radius of the corresponding inscribed circle).\footnote{The skeleton together with the corresponding radius function constitute the medial axis transform.} }

\begin{figure}
\includegraphics[width=1\columnwidth,height=0.76\columnwidth]{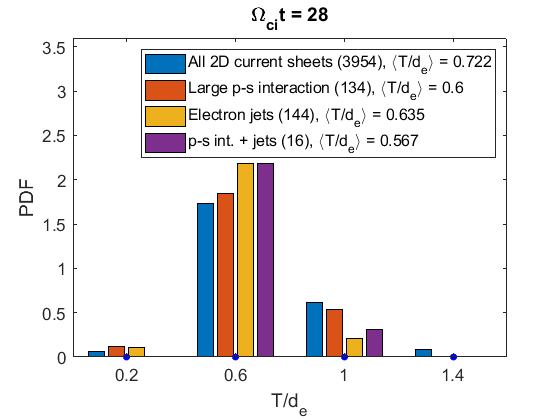}
\includegraphics[width=1\columnwidth,height=0.76\columnwidth]{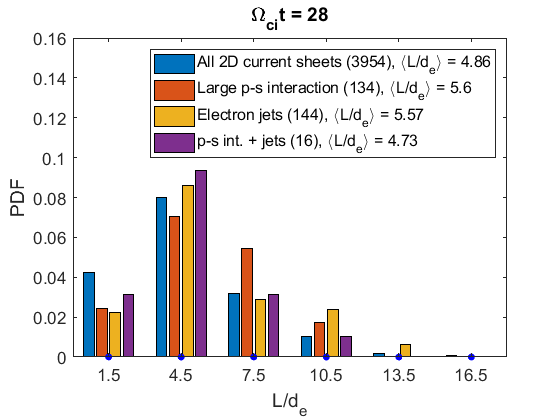}
\includegraphics[width=1\columnwidth,height=0.76\columnwidth]{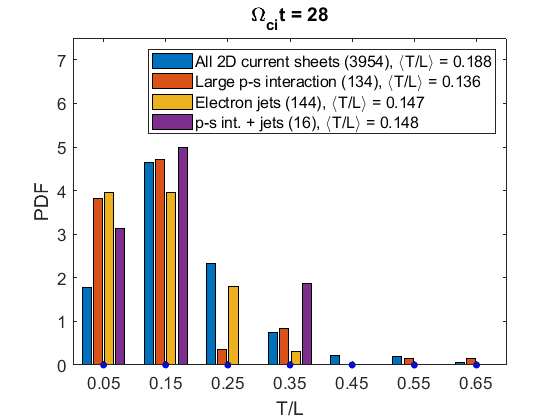}
\caption{Histograms of half-thicknesses $T$ (top),  half-lengths $L$ (middle), and aspect ratios $T/L$ (bottom) of the 2D current sheets, with different sets shown in different colors. The number of 2D current sheets in each case is shown in parentheses. The blue dots mark the centers of the bins. The data denoted by different colors belong to the same bins but are shifted for clarity.}
\label{sizes}
\end{figure}

The medial axis of each 2D current sheet and the distance of each point on it to the boundary (the radius function) were obtained using the Matlab Image Processing Toolbox functions ``bwskel'' and ``bwdist'', respectively. For each current sheet, the half-thickness ($T$) is then estimated by averaging the radius function over the points of the skeleton. We define the half-length ($L$) of a current sheet as half the largest distance between two points belonging to the sheet structure. The distributions of the thicknesses and lengths of the identified current sheets are presented in Figure \ref{sizes}. The figure also shows the distribution of the aspect ratios of the current sheets. The distributions show that these are electron-scale current sheets, as expected. \footnote{It should be noted that the lack of ion scale current sheets is due to the choice of energy injection scale and box size. However, the relevant observation is that kinetic scale turbulence indeed creates electron scale current sheets.}

Figure~\ref{sizes} also contains a similar analysis performed for two specific subsets of the 2D current sheets. The first subset is characterized by large values of the so-called pressure-strain interaction, which may indicate strong energy dissipation. The second subset is characterized by large changes in the velocity of the electron flow, which may indicate the presence of outflows expected in electron-only reconnection. The distributions for both subsets generally overlap. The intersection of these two subsets, shown in purple, seems to follow a similar trend. However, since it contains only 16 current sheets, it has limited statistical significance.

We define the pressure-strain interaction density as $\rate_\alpha = -(\mathbf{P}_\alpha\cdot\nabla)\cdot\mathbf{U}_\alpha$, which is related to the global rate of change of the thermal energy $\cal E_\alpha$~\citep{Yang2017,yang2022}:
\begin{eqnarray}
    \langle\partial_t\cal E_\alpha \rangle=-\langle(\mathbf{P}_\alpha\cdot\nabla)\cdot\mathbf{U}_\alpha\rangle = \langle \rate_\alpha \rangle.
    \label{eq:dis_rate}
\end{eqnarray}
Here,  $\alpha$ refers to the particle species, $\mathbf{P}_\alpha$ is the pressure tensor, $\mathbf{U}_\alpha$ is the fluid velocity field, and $\langle...\rangle$ represents the average over the whole domain. Periodic boundary conditions are assumed in the derivations of this equation. We place a 2D current sheet into the large-pressure-strain interaction subset if there exists a point within a window of $2.5d_e\times2.5d_e$ centered on its skeleton where $\rate = \sum_\alpha R_\alpha$ is at least ten times its root-mean-square (rms) value over the whole domain. The window size was chosen to pick up peaks in $\mathcal{R}$, similar to the ones seen in the top-right panel of Figure~\ref{reconnection}. Such peaks are typically found around a current-sheet skeleton but they do not overlap with it.

Similarly, we place a 2D current sheet into the large-electron-outflow subset if it contains a large variation of the electron velocity ${\bf U}_e$ in the vicinity of its skeleton.  Specifically, we look for variations in the in-plane electron fluid velocity larger than 95\% of the electron Alfvén velocity $V_{Ae}$ within $5d_e\times5d_e$ windows centered on the skeletons. (Here $V_{Ae}$ is defined with the rms value of the in-plane magnetic field.) Such structures are of interest because the presence of electron outflows is one of the signatures (but not a guarantee, see below) of magnetic reconnection, and also because velocity gradients explicitly enter the expression for $\rate$, Equation~\ref{eq:dis_rate}. {The distributions of current sheet sizes corresponding to different subsets look similar and have comparable averages. However, more intense energy dissipation and/or electron outflows seem to favor more anisotropic current sheets.}  

\subsection{Intermittency of energy dissipation}

\begin{figure}
\includegraphics[width=\columnwidth]{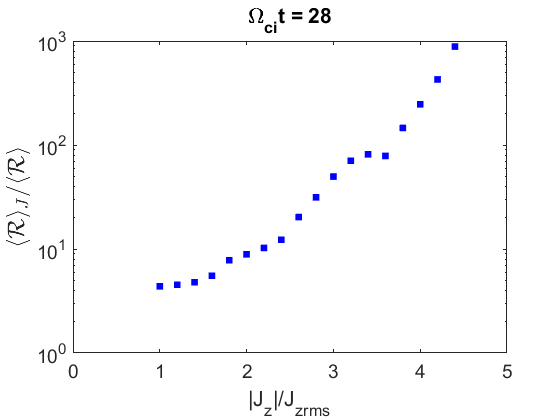}
\includegraphics[width=\columnwidth]{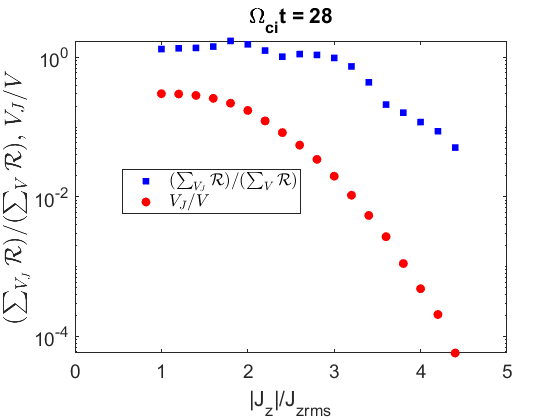}
\caption{Top: Average pressure-strain interaction around points of the current sheet skeletons where the current density is above the threshold shown on the horizontal axis. The pressure-strain interaction is normalized to its average over the whole domain and the current density is normalized to its rms value. Bottom: Fraction of the overall pressure-strain interaction contributed by the vicinity of points of current sheet skeletons (in blue). The fractional volume occupied by the region around skeletons where pressure-strain interaction was computed (in red).}
\label{dissipation}
\end{figure}

In order to illustrate the intermittency of energy dissipation (i.e., a significant contribution to the dissipation coming from a small fraction of the domain), we consider the relation between current density and energy dissipation in the 2D current sheets. {The top panel of Figure~\ref{dissipation} shows the average $\mathcal{R}$ computed within $2.5d_e \times 2.5d_e$ windows centered on those points of current sheet skeletons where the current density is above a given threshold.} This figure demonstrates that, indeed, the average $\mathcal{R}$ is higher in the vicinity of intense current sheets, reaching values that are over a hundred times the average where the current densities are four times the rms.

The bottom panel of Figure~\ref{dissipation} shows, in blue, $\mathcal{R}$ integrated over the corresponding vicinities of the skeletons, normalized to $\mathcal{R}$ integrated over the whole domain. In red, we show the corresponding areas (2D volumes) over which $\mathcal{R}$ was integrated, normalized to the volume of the whole domain. This plot illustrates the intermittency of intense energy dissipation. Notably, $\mathcal{R}$ computed in regions where the current density exceeds twice its rms value, adds up to more than the total $\mathcal{R}$ over the whole domain while occupying about 20\% of the volume of the whole domain (this is possible due to partial cancellation with sites where $\mathcal{R}<0$). Thus, energy dissipation in this low $\beta_e$ turbulent environment is seen to be strongly intermittent, similar to what has been observed in simulations of other collisionless plasma environments~\citep[e.g.,][]{wan2012,wan2015,camporeale2018}.

\subsection{Electron-only reconnection}\label{Rec}

\begin{figure*}
\centering
\includegraphics[width=\columnwidth,height=0.8\columnwidth]{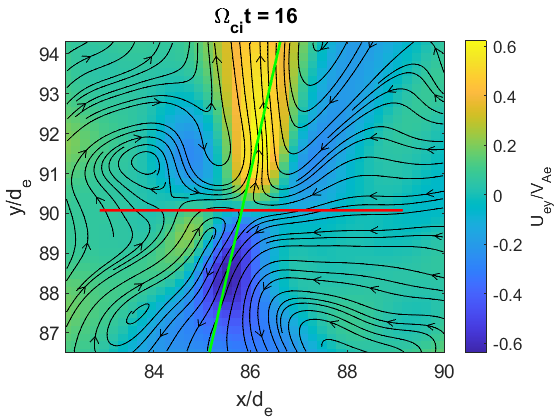}\hfill
\includegraphics[width=\columnwidth,height=0.8\columnwidth]{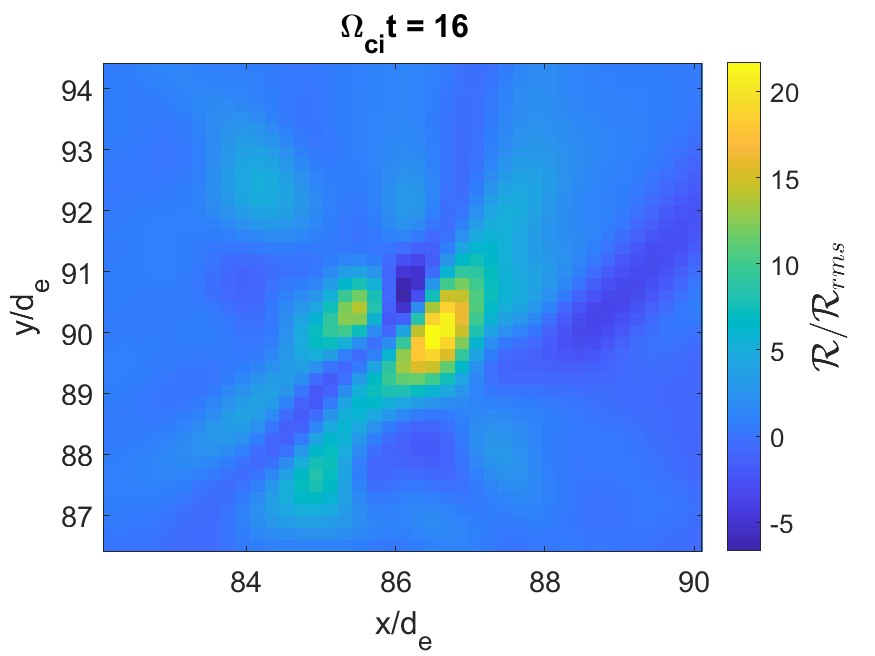}\\
\includegraphics[width=\columnwidth,height=0.8\columnwidth]{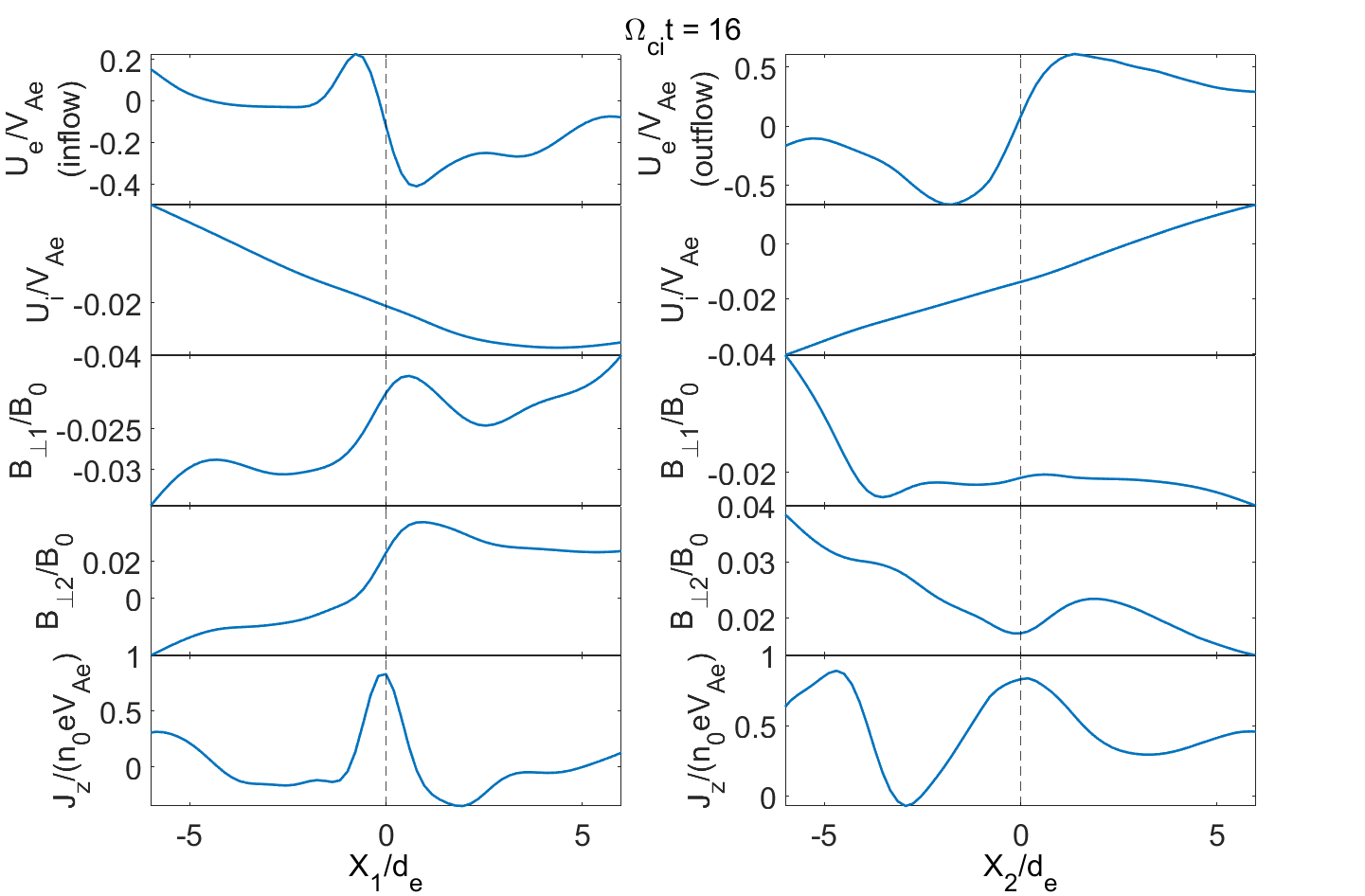}\hfill
\includegraphics[width=\columnwidth,height=0.8\columnwidth]{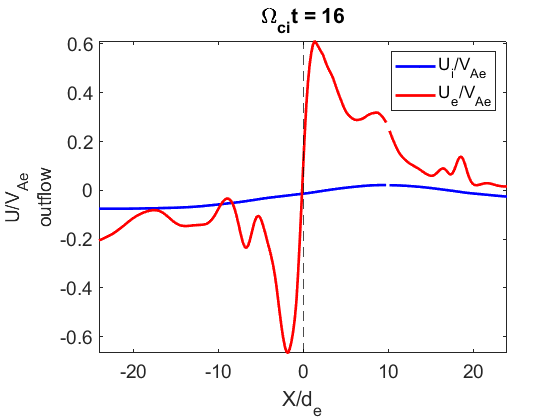}
\caption{Example of reconnection candidate. Top-left: The color map shows the y component of the electron flux and the arrows show the electron fluid velocity. Bottom-left: Profiles of different fields along the inflow (direction $X_1$; red line in top panel) and the outflow (direction $X_2$; green line in top panel). Top-right: Color map of pressure-strain interaction. Bottom-right: Ion (blue) and electron (red) velocities along outflow. The horizontal scale spans about 5$d_i$, or half the domain length in the field-perpendicular plane. There is no ion coupling to the electron motion.}
\label{reconnection}
\end{figure*}

Among the analyzed 3954 2D current structures on 63 planes, we have found 16 sites that simultaneously exhibit large $\rate$ and nearly electron-Alfv\'enic jumps in the electron flow velocity, belonging to about seven 3D structures. We have observed that the presence of large velocity jumps on electron scales, $\Delta U_e \sim V_{Ae}$, does not yet guarantee that the corresponding current sheets are undergoing magnetic reconnection. Indeed, manual inspection of the seven candidates with large $\Delta U_e$ and $\rate$ yielded just five 3D structures (representing about 0.3\% of the 3954 2D current sheets) exhibiting magnetic configuration and other signatures typical of reconnecting current sheets. The same analysis done at $\Omega_{ci}t=16$ showed that about 2\% of the 2D current sheets at that time display signatures of magnetic reconnection. One of such structures, shown in Figure~\ref{reconnection}, exhibits the largest value of $\rate$ in the entire domain at either $\Omega_{ci}t=16$ or $\Omega_{ci}t=28$ (about 22 times the rms value at $\Omega_{ci}t=16$). 

The top-left panel in Figure~\ref{reconnection} shows the electron fluid velocity $U_{ey}$ normalized to the electron Alfvén speed, with the in-plane electron streamlines in black. Inflow and outflow patterns are clearly present. The corresponding profiles of the in-plane electron and ion fluid velocities, the in-plane magnetic field, and the out-of-plane current density along the inflow and outflow directions are shown in the bottom-left panel of Figure~\ref{reconnection}.  The direction of the outflow cut (marked by the green line) was chosen so that it connects the maxima of the $y$-component of the electron fluid velocity. The inflow direction (red line) was chosen along the x-direction. All the fluid velocities are normalized to the electron Alfvén speed. We note that, unlike the case of 2D reconnection, nearly Alfvénic velocities are seen not only along the outflow but also along the inflow, {suggesting that there is little magnetic to bulk motion energy conversion in this reconnection event.} Similar enhanced inflows were also observed in 3D reconnection simulations  in~\cite{pyakurel2021}, where it was attributed to a significant mass outflow along the background field.

The top-right panel of Figure~\ref{reconnection} shows the large peak in $\rate$ that was found around the reconnection site, which corresponds to the most intense dissipation event found in the entire domain at this particular time slice. The bottom right panel displays the ion and electron outflows in an enlarged region spanning half the field-perpendicular domain. As can be seen in the figure, ion flows are significant (in relation to the ion Alfv\'en speed) and do exhibit a reversal around the reconnection site. At the same time, ion and electron motions remain decoupled until outflows begin to interact with other turbulent structures on the scales comparable to ion gyroradius (recall that here $\beta_i \approx 1$ and hence $\rho_i \approx d_i \approx 10 d_e$). In this sense, it is appropriate to interpret the presented reconnection event as electron-scale reconnection. The reconnection site shown in Figure~\ref{reconnection} is part of a 3D current structure elongated along the background field, with a field-parallel length of approximately 300$d_e$. Its average half-thickness and half-length are $0.77d_e$ and $7.5d_e$, respectively. {To estimate the value of $\mathcal{R}$ associated to this 3D structure, we consider  $2.5d_e\times2.5d_e$ windows centered on the 2D current-sheet skeletons that span the 3D structure over 30 field-perpendicular planes, and we compute $\mathcal{R}$ within these windows. $\mathcal{R}$ computed this way, for this particular 3D structure, turns out to contribute 25\% of the total pressure-strain interaction} in the whole domain, in only 0.26\% of the total volume.

We have also examined other metrics typically associated with reconnection for the structure discussed above. It was found that this structure corresponds to one of the largest values of magnetic shear in the simulation, which was analyzed by studying the mapping of the field lines between two $xy$ planes of the simulation located at different $z$~\citep[e.g.,][and references therein]{Daughton2014}. It also corresponds to large electron agyrotropy~\citep{Scudder2008}, which signals departures of the electron velocity distribution function from cylindrical symmetry about the local magnetic field.

\section{Discussion}
\label{conclusions}

In this work, we studied 3D kinetic-scale turbulence in a plasma with low electron beta, $\beta_e=0.1$ and $\beta_i=1$, a  regime that may be encountered in the vicinity of the solar corona~\cite[e.g.,][]{shi2023}, in the Earth's magnetosheath~\cite[e.g.,][]{chen_boldyrev2017}, and in other astrophysical environments where electrons are colder than ions. Recent observational, analytical, and numerical studies showed that turbulent cascade in such a plasma can reach scales smaller than the electron inertial scale $d_e$, essentially decoupling from the ion dynamics \cite[e.g.,][]{chen_boldyrev2017,roytershteyn2019,vega2020}. The resulting kinetic turbulence may form electron-scale current sheets and may trigger the so-called electron-only reconnection events, where electrons remain decoupled from the ions due to the relatively small spatial extent of the reconnection region. 

The simulation analyzed in this work was performed using the SPS code, which utilizes an algorithm based on the spectral decomposition of the distribution function. Such a technique occupies an intermediate space between fluid and fully kinetic methods traditionally used in studies of kinetic-scale turbulence and potentially offers a combination of properties (accuracy, implicit time-stepping, and the lack of {\em a priori} assumption on the ordering of various terms) that may be particularly advantageous for studies of turbulence.  
 
The 3D simulation demonstrated that for the plasma parameters and for the relatively weak magnetic perturbations considered $\delta B/B_0 \sim 0.1 $, the turbulence does generate electron-scale current sheets and that some of them appear to undergo magnetic reconnection without efficient coupling to the ions. Statistical analysis of the geometrical properties of the detected current sheets was performed using an algorithm based on the medial axis transform. In particular, a typical half-thickness of the current sheets was found to be on the order of $d_e$ or below, while their half-length falls between the $d_e$ and $d_i$ scales (recall that in order to couple to the ion dynamics, the current sheet length needs to significantly exceed $d_i$ \cite[e.g.,][]{pyakurel2019,vega2020,stawarz2022}). 

The pressure-strain interaction, used as a measure of energy dissipation, exhibited high intermittency, with most of the total dissipation originating from current sheets that occupied merely 20\% of the volume. This finding aligns with previous studies that have also demonstrated strong intermittency in energy dissipation in various plasma regimes~\citep[e.g.,][]{wan2012,wan2015,camporeale2018}, highlighting the ubiquity of intermittent structures and dissipation in turbulent plasmas. Some of the current sheets exhibiting the largest peaks in the pressure-strain interaction were found to be associated with Alfv\'enic electron jets and magnetic configurations typical of reconnection (i.e., there is an X-point in the magnetic field in the plane perpendicular to the local field). These reconnection candidates represent about $1$\% of all the current sheets identified. Among such structures, the one associated with the largest peak in the value of the pressure-strain interaction was observed to contribute approximately 25\% of the total pressure-strain interaction in the whole domain, while occupying only 0.26\% of the total volume.

\section*{Acknowledgements}
This research was partially supported by the Los Alamos National Laboratory (LANL) through its Center for Space and Earth Science (CSES). CSES is funded by LANL's Laboratory Directed Research and Development (LDRD) program under project number 20210528CR. The work of SB was partly supported by NSF grant no. PHY-2010098 and by DOE grant No. DE-SC0018266. VR was partly supported by NASA grant 80NSSC21K1692 and by DOE grant DE-SC0019315. Computational resources were provided by the NSF ACCESS program (Allocations No. TG-PHY110016 and TG-ATM180015) and by the NASA High-End Computing Program through the NASA Advanced Supercomputing Division at Ames Research Center. 

\section*{Data Availability}

The simulation data used in this study is available at Harvard Dataverse \url{https://dataverse.harvard.edu/privateurl.xhtml?token=5b7a27d7-489d-4dd2-ad16-adbd6caa1fe2}.



\bibliographystyle{mnras}
\bibliography{electron_refs} 








\bsp	
\label{lastpage}
\end{document}